\documentclass[acmtog,timestamp]{acmart}
\usepackage[normalem]{ulem}
\usepackage{todonotes}
\usepackage{subcaption}
\AtBeginDocument{%
  \providecommand\BibTeX{{%
    \normalfont B\kern-0.5em{\scshape i\kern-0.25em b}\kern-0.8em\TeX}}}

\newif \ifmakeAnon
\makeAnontrue
\usepackage{censor}
\ifmakeAnon \else \StopCensoring \fi

\setcopyright{acmcopyright}
\copyrightyear{2020}
\acmYear{2020}
\acmDOI{10.1145/1122445.1122456}

\acmSubmissionID{papers\_262}
\acmJournal{TOG}
\acmVolume{39}
\acmNumber{4}
\acmArticle{1}
\acmYear{2020}
\acmMonth{7}
\acmDOI{10.1145/3386569.3392413}

\newcommand{\npapers}{374}
\newcommand{\ncodes}{151}
\newcommand{\npapersIV}{127}
\newcommand{\npapersVI}{119}
\newcommand{\npapersVIII}{128}
\newcommand{\ncodesIV}{37}
\newcommand{\ncodesVI}{47}
\newcommand{\ncodesVIII}{67}
\newcommand{\pccodesIV}{29.1}
\newcommand{\pccodesVI}{39.5}
\newcommand{\pccodesVIII}{52.3}
\newcommand{\pcodesIVVI}{0.13}
\newcommand{\pcodesVIVIII}{0.086}
\newcommand{\pcodesIVVIII}{1.3\,{10}^{-3}}

\newcommand{\pccodesacad}{45.4}
\newcommand{\pccodesindus}{31.3}
\newcommand{\pcodesindus}{0.031}
\newcommand{\preplyear}{0.27}
\newcommand{\preplreviewer}{0.13}
\newcommand{\nmodifiedcodes}{68}
\newcommand{\nhardmodifiedcodes}{20}
\newcommand{\nlongcodes}{27}
\newcommand{\ncodesdeep}{25}
\newcommand{\npapersdeep}{56}
\newcommand{\ncodesnodeep}{126}
\newcommand{\npapersnotdeep}{318}
\newcommand{\fcodesdeep}{44.6}
\newcommand{\fcodesnodeep}{39.6}
\newcommand{\pcodesdeepn}{0.48}
\newcommand{\narxiv}{27}
\newcommand{\nopenaccess}{44}
\newcommand{\nsourcecode}{133}
\newcommand{\nbinaries}{18}
\newcommand{\citcodeIV}{67}
\newcommand{\citnocodeIV}{43}
\newcommand{\pcitcodenocodeIV}{0.045}
\newcommand{\nhardissues}{5}
\newcommand{\nsoftissues}{19}
\newcommand{\nunspeclicence}{60}
\newcommand{\nodoc}{11}

\newcommand{\doczeroIV}{6}
\newcommand{\doczeroVI}{16}
\newcommand{\doczeroVIII}{14}
\newcommand{\doczeroTot}{36}
\newcommand{\deepIV}{4}
\newcommand{\deepVI}{4}
\newcommand{\deepVIII}{17}
\newcommand{\deepTot}{25}
\newcommand{\nolicIV}{16}
\newcommand{\nolicVI}{24}
\newcommand{\nolicVIII}{20}
\newcommand{\nolicTot}{60}
\newcommand{\dataIV}{5}
\newcommand{\dataVI}{7}
\newcommand{\dataVIII}{18}
\newcommand{\dataTot}{30}
\newcommand{\pccodesGeo}{51.9}
\newcommand{\pccodesIma}{57.9}
\newcommand{\pccodesRen}{47.9}
\newcommand{\pccodesAni}{26.9}
\newcommand{\pccodesFab}{17.1}
\newcommand{\pccodesVR}{31.8}

\begin{document}

\title{Code Replicability in Computer Graphics}


\author{Nicolas Bonneel}
\email{nicolas.bonneel@liris.cnrs.fr}
\affiliation{
  \institution{Univ Lyon, CNRS}
  \orcid{0000-0001-5243-4810}
  \city{Lyon}
  \country{France}
}

\author{David Coeurjolly}
\email{david.coeurjolly@liris.cnrs.fr}
\affiliation{
  \institution{Univ Lyon, CNRS}
  \orcid{0000-0003-3164-8697}
  \city{Lyon}
  \country{France}
  }   
\author{Julie Digne}
\email{julie.digne@liris.cnrs.fr}
\affiliation{
  \institution{Univ Lyon, CNRS}
  \orcid{0000-0003-0905-0840}
  \city{Lyon}
  \country{France}}

\author{Nicolas Mellado}
\email{nicolas.mellado@irit.fr}
\affiliation{
  \institution{Univ Toulouse, CNRS}
  \orcid{0000-0003-2180-4318}
  \city{Toulouse}
  \country{France}
}

\begin{abstract}
	Being able to duplicate published research results is an
        important process of conducting research whether to build
        upon these findings or to compare with them.  This process is
        called ``replicability'' when using the original authors'
        artifacts (e.g., code), or ``reproducibility'' otherwise
        (e.g., re-implementing algorithms).  Reproducibility and
        replicability of research results have gained a lot of
        interest recently with assessment studies being led in various
        fields, and they are often seen as a trigger for better result
        diffusion and transparency.  In this work, we assess
        replicability in Computer Graphics, by evaluating whether the code is available and whether it works properly.
	As a proxy for this field we
        compiled, ran and analyzed \ncodes~codes out of
        \npapers~papers from 2014, 2016 and 2018 SIGGRAPH
        conferences. This analysis shows a clear increase in the
        number of papers with available and operational research codes
        with a dependency on the subfields, and indicates a
        correlation between code replicability and citation count. We
        further provide an interactive tool to explore our results and
        evaluation data.
\end{abstract}



\begin{teaserfigure}
\includegraphics[width=1.0\textwidth]{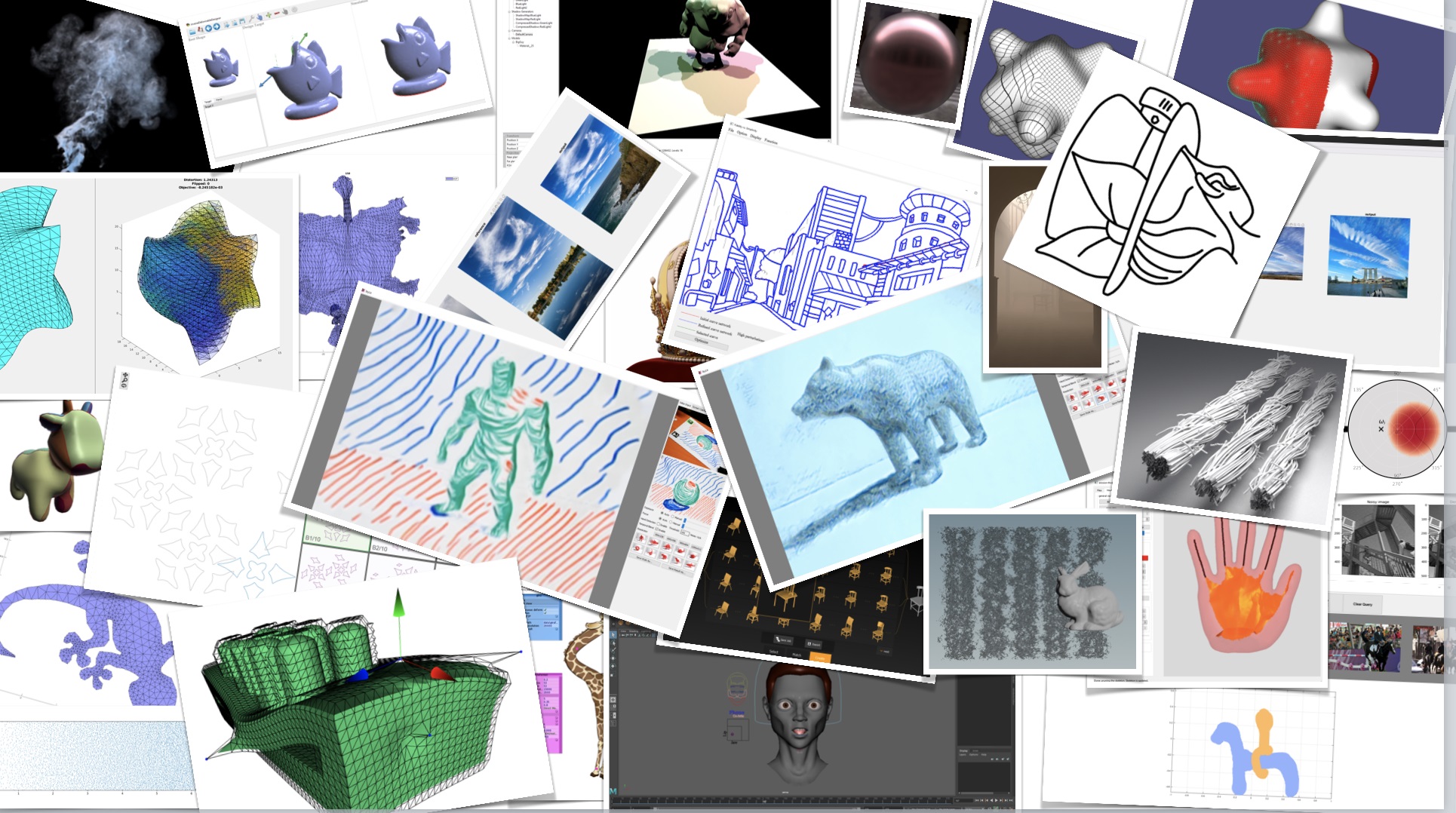}
\caption{We ran \ncodes~codes provided by papers published at SIGGRAPH 2014, 2016 and 2018.
We analyzed whether these codes could still be run as of 2020 to provide a replicability score,
and performed statistical analysis on code sharing. \footnotesize{Image credits: Umberto Salvagnin, \_Bluenose Girl, Dimitry B., motiqua, Ernest McGray Jr., Yagiz Aksoy, Hillebrand Steve. 3D models by \href{http://www.loramel.net/}{Martin Lubich} and Wig42.} }
\end{teaserfigure}
\maketitle

\section{Introduction}
The ability to reproduce an experiment and validate its results is a cornerstone of scientific research, a key to our understanding of the world.
Scientific advances often provide useful tools, and build upon a vast body of previous work
published in the literature. As such, research that cannot be
reproduced by peers despite best efforts often has limited value,
  and thus impact, as it
does not benefit to others, cannot be used as a basis for further
research, and casts doubt on published results. Reproducibility is
also important for comparison purposes since new methods are often
seen in the light of results obtained by published competing approaches.
Recently serious concerns have emerged in various scientific communities
from psychological sciences~\cite{open2015estimating} to artificial
intelligence~\cite{hutson2018artificial} over the lack of
reproducibility, and one could wonder about the state of computer
graphics research in this matter.

In the recent trend of open science and reproducible research, this paper aims at assessing the state of replicability of papers published at ACM Transactions on Graphics as part of SIGGRAPH conferences. Contrary to reproducibility which assesses how results can be obtained by independently reimplementing published papers -- an overwhelming task given the hundred papers accepted yearly to this event -- replicability ensures the authors' own codes run and produce the published results. While sharing code is not the only available option to guarantee that published results can be duplicated by a practitioner -- after all, many contributions can be reimplemented from published equations or algorithm descriptions with more or less effort -- it remains an important tool that reduces the time spent in reimplementation, in particular as computer graphics algorithms get more sophisticated.

Our contributions are twofold. First, we analyze code sharing practices and replicability in computer graphics.
We hypothesize strong influence of topics, an increase of replicability over time similar to the trend observed in artificial intelligence~\cite{hutson2018artificial}, and an increased impact of replicable papers, as observed in image processing~\cite{Vandewalle19}. To evaluate these hypotheses, we manually collected source codes of SIGGRAPH 2014, 2016 and 2018 papers and ran them, and when possible, assessed how they could replicate results shown in the paper or produce reasonably similar results on different inputs.
Second, we provide detailed step-by-step  instructions to make these software packages run (in practice, in many cases, code adaptations had to be done due to dependencies having evolved) through a website, thus becoming a large code review covering \ncodes~codes obtained from \npapers~SIGGRAPH papers. We hope this platform can be used collaboratively in the future to help researchers having difficulties reproducing published results.

Our study shows that:
\begin{itemize}
\item Code sharing is correlated with paper citation count, and has improved over time.
\item Code sharing practices largely vary with sub-fields of computer graphics.
\item It is often not enough to share code for a paper to be replicable. Build instructions with precise dependencies version numbers as well as example command lines and data are important.
\end{itemize}

\section{Prior work}
\label{sec:prior}

The impact of research involves a number of parameters that are
independent of the quality of the research itself, but of practices
surrounding it. Has the peer review process been fairly conducted?
Are the findings replicable? Is the paper accessible to the citizen? A number of these questions have been studied in the past within various scientific communities, which this section reviews.

\paragraph{\textbf{Definitions.}}
Reproducible research has been initiated in computer science~\cite{claerbout1992electronic} via the automation of figures production within scientific articles. Definitions have evolved~\cite{plesser2018reproducibility} and have been debated~\cite{goodman2016does}. As per ACM standards~\cite{acmrepl}, \textit{repeatability} indicates the original authors can duplicate their own work, \textit{replicability} involves other researchers duplicating results using the original artifacts (e.g., code) and hardware, and \textit{reproducibility} corresponds to other researchers duplicating  results with their own artifacts and hardware -- we will hence use this definition. We however mention that various actors of replicable research have advocated for the \emph{opposite} definition: replicability being about answering the same research question with new materials while reproducibility involves the original artifacts~\cite{barba2018terminologies} -- a definition championed by the National Academies of Sciences~\shortcite{national2019reproducibility}.

\paragraph{\textbf{Reproducibility and replicability in experimental sciences.}}
Concerns over lack of reproducibility have started to emerge in
several fields of studies, which has led to the term ``reproducibility
crisis''~\cite{pashler2012replicability}.  In experimental
  sciences, replicability studies evaluate
  whether claimed hypotheses are validated from
  observations ({e.g.,} whether the null hypothesis is consistently rejected
  and whether effect sizes are similar). In different fields of
psychology and social sciences, estimations of replication
rates have varied between 36\% out of 97 studies with significant
results, with half the original effect size~\cite{open2015estimating},
50\%-54\% out of 28 studies~\cite{klein2018many}, 62\% out of 21
Nature and Science studies with half the original effect
size~\cite{camerer2018evaluating}, and up to roughly 79\% out of 342
studies
~\cite{makel2012replications}. In oncology, a reproducibility rate of 11\% out of 53 oncology papers has been estimated~\cite{begley2012drug}, and a collaboration between Science Exchange and the Center for Open Science (initially) planned to replicate 50 cancer biology studies~\cite{baker2017cancer}. Over 156 medical studies reported in newspapers, about 49\% were confirmed by meta-analyses~\cite{dumas2017poor}.
A survey published in Nature~\cite{baker20161} showed large disparities among scientific fields: respondents working in engineering believed an average of 55\% of published results are reproducible ($N=66$), while in physics an average of 73\% of published results were deemed reproducible ($N=91$).

This has resulted in various debates and solutions such as reducing hypothesis testing acceptance thresholds to $p<0.005$~\cite{benjamin2018redefine} or simply abandoning hypothesis testing and p-values as binary indicators~\cite{mcshane2019abandon}, providing confidence intervals and using visualization techniques~\cite{cohen2016earth}, or improving experimental protocols~\cite{begley2013reproducibility}.

While computer graphics papers occasionally include experiments such as perceptual user studies, our paper focuses on code replicability.

\paragraph{\textbf{Reproducibility and replicability in computational sciences.}}
In hydrology, Stagge et al.~\shortcite{stagge2019assessing} estimate via a
survey tool that 0.6\% to 6.8\% of 1,989 articles (95\% Confidence Interval) can be reproduced using the available data, software and code -- a major reported issue being the lack of directions to use the available artifacts (for 89\% of tested articles).
High energy physicists, who depend on costly, often unique, experimental setups (e.g., the Large Hadron Collider) and produce enormous datasets, face reproducibility challenges both in data collection and processing~\cite{chen2019open}. Such challenges are tackled by rigorous internal review processes before data and tools are opened to larger audiences. It is argued that analyses should be automated from inception and not as an afterthought.
Closer to our community is the replication crisis reported in artificial intelligence~\cite{hutson2018artificial,gundersen2018state}. Notably, the authors surveyed 400 papers from top AI conferences IJCAI and AAAI, and found that 6\% of presenters shared their algorithm's code, 54\% shared pseudo-code, 56\% shared their training data, and 30\% shared their test data, while the trend was improving over time.
In a recent study on the reproducibility of IEEE Transactions on Image Processing papers~\cite{Vandewalle19}, the authors showed that, on average, code availability approximately doubled the number of citations of published papers.
Contrary to these approaches, we not only check for code availability, but also evaluate whether the code compiles and produces similar results as those found in the paper, with reasonable efforts to adapt and debug codes when needed.

\sloppy Efforts to improve reproducibility are nevertheless flourishing from early recommendations such as building papers using \texttt{Makefiles} in charge of reproducing figures~\cite{schwab2000making} to various reproducibility badges proposed by ACM~\cite{acmrepl} in collaboration with the \textit{Graphics Replicability Stamp Initiative}~\cite{grsi2016}. Colom et al. list a number of platforms and tools that help in reproducible research~\shortcite{colom2018overview}. Close to the interest of the computer graphics community, they bring forward the IPOL journal~\cite{colom2015ipol} whose aim is to publish image processing codes via a web interface that allows to visualize results, along with a complete and detailed peer-reviewed description of the algorithm. They further mention an initiative by GitHub~\shortcite{github2016} to replicate published research, though it has seen very limited success (three replications were initiated over the past three years).
In Pattern Recognition, reproducible research is awarded with the
  Reproducible Label in Pattern
  Recognition organized by the biennal Workshop on Reproducible Research in Pattern Recognition~\cite{kerautret2017reproducible,kerautret2019reproducible}.
Programming languages and software engineering communities have created the Artifact Evaluation Committees for accepted papers~\cite{AEC}, with incentives such as rewarding with additional presentation time at the conference and an extra page in the proceedings, with special recognition for best efforts.

Other initiatives include reproducibility challenges such as the one organized yearly since 2018 by the ICLR conference in machine learning~\cite{Pineau2019iclr} that accepts submissions aiming at reproducing published research at ICLR. In 2018, reproducibility reports of 26 ICLR papers were submitted, out of which 4 were published in the ReScience C journal.

\paragraph{\textbf{Open access}}
Software bugs have had important repercussions on collected data and analyses, hence pushing for open sourcing data and code. Popular examples include Microsoft Excel that converts gene names such as SEPT2 (for Septin 2) to dates~\cite{ziemann2016gene}, or a bug in widely used fMRI software packages that resulted in largely inflated false-positive rates, possibly affecting many published results~\cite{eklund2016cluster}.
Recently, Nature Research has enforced an open data policy~\cite{natcom2018}, stated in their policies as \textit{authors are required to make materials, data, code, and associated protocols promptly available to readers without undue qualifications}, and proposes a journal focused on sharing high re-use value data called \textit{Scientific Data}~\cite{scidata}.
Other platforms for sharing scientific data include the \textit{Open Science Framework}~\cite{osf}. Related to code, Colom et al.~\shortcite{colom2018overview} reports the websites \textit{mloss} that lists machine learning codes, \textit{RunMyCode} for scientists to share code associated with their research paper, or \textit{ResearchCompendia} that stores data and codes.
Long-term code availability is also an issue, since authors' webpages are likely to move according to institution affiliations so that code might be simply unavailable. Code shared on platforms such as GitHub is only available as long as the company {exists} which can also be an issue, if limited. For long-term code storage, the Software Heritage initiative~\cite{dicosmo17} aims at crawling the web and platforms such as GitHub, Bitbucket, Google code etc. for open source software and stores them in a durable way.
Recently, the Github Archive Program~\cite{github2020archiveprogram} pushed these ideas further and propose a pace layer strategy where code is archived at different frequencies (real-time, monthly, every 5 years), with advertised lifespans up to 500 years and possibly 10,000 years.

\paragraph{\textbf{Other assessments of research practices}} Reproducibility of paper acceptance outcome has been assessed in machine learning. In 2014, the prestigious NIPS conference (now NeurIPS) has performed the \textit{NIPS consistency experiment}: a subset of 170 out of 1678 submissions were assigned to two independent sets of reviewers, and consistency between reviews and outcomes were evaluated. The entire process, results, and analyses,  were shared on an open platform~\cite{nips2014}. Decisions were inconsistent for 43 out of 166 reviewed papers (4 were withdrawn, 101 were rejected by both committees, 22 were accepted by both committees). Other initiatives for more transparent processes include the sharing of peer reviews of published papers on platforms such as \textit{OpenReview}~\cite{openreview} or directly by journals~\cite{royal}, and the post-publication monitoring for misconducts or retractions on platforms such as PubPeer and RetractionWatch~\cite{didier2018new}.

\section{Method}

Our goal is to assess trends in replicability in computer graphics. We chose to focus on the conference in the field with highest exposure, ACM SIGGRAPH, as an upper bound proxy for replicability. Although this hypothesis remains to be verified, this conference more often publishes completed research projects
 as opposed to preliminary exploratory ideas that are more often seen in smaller venues which could explain lower code dissemination.
To estimate a trend over time, we focus on three SIGGRAPH conferences: SIGGRAPH 2014 (Vancouver, 127 accepted papers), 2016 (Anaheim, 119 accepted papers), and 2018 (Vancouver, 128 accepted papers). We did not include SIGGRAPH 2019 (Los Angeles) since authors sometimes need time to clean up and publish their code after publication. We did not include SIGGRAPH Asia nor papers published in ACM Transactions on Graphics outside of the conference main track to reduce variability in results and keep a more focused scope. We chose a two-year interval between conferences in the hope to get clearer trends, and to keep a tractable number of papers to evaluate.

We searched for source codes as well as closed-source binaries 
for all papers. We restricted our search to original implementations and reimplementations authored and released by the original authors of the paper, excluding reimplementations by others, as we aim at assessing replicability and not reproducibility (see Sec.~\ref{sec:prior}). For each paper, we report the objective and subjective information described below.

\begin{figure}[!tb]
\centering
\begin{subfigure}{.45\columnwidth}
 \centering
 \includegraphics[width=\columnwidth]{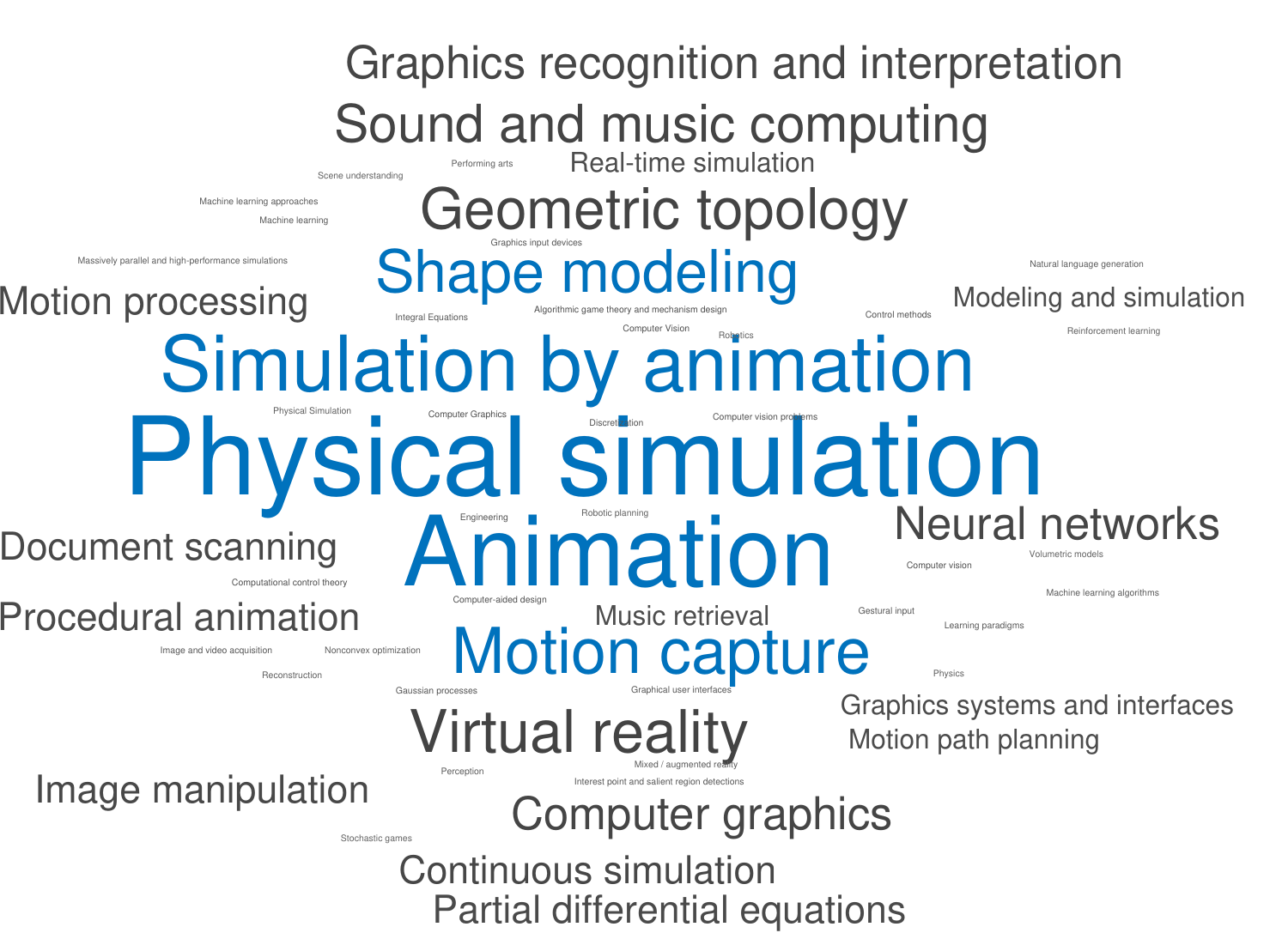}
 \caption{Animation}
\end{subfigure}%
\begin{subfigure}{.45\columnwidth}
 \centering
 \includegraphics[width=\columnwidth]{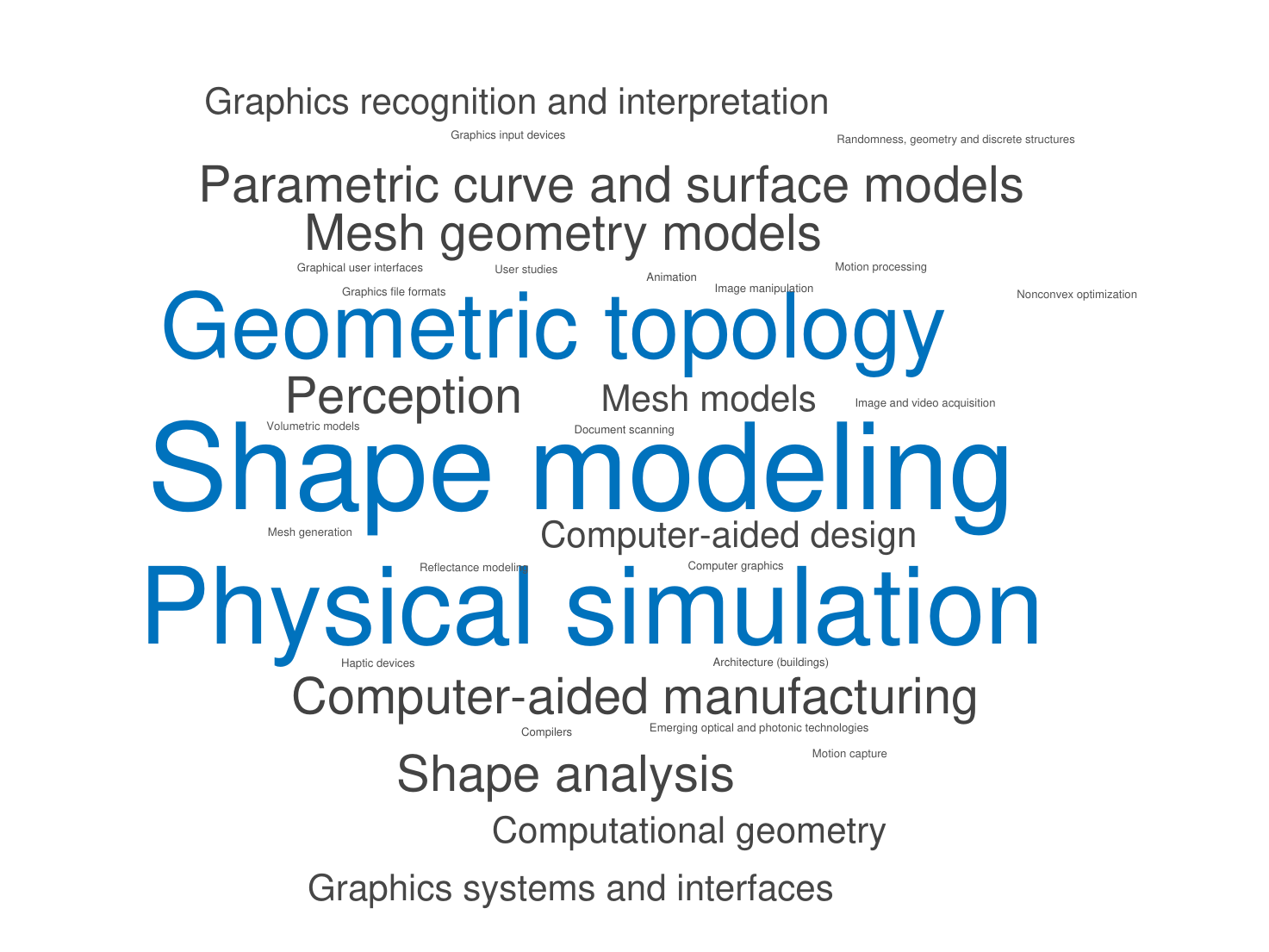}
 \caption{Fabrication}
\end{subfigure}%
\newline
\begin{subfigure}{.45\columnwidth}
 \centering
 \includegraphics[width=\columnwidth]{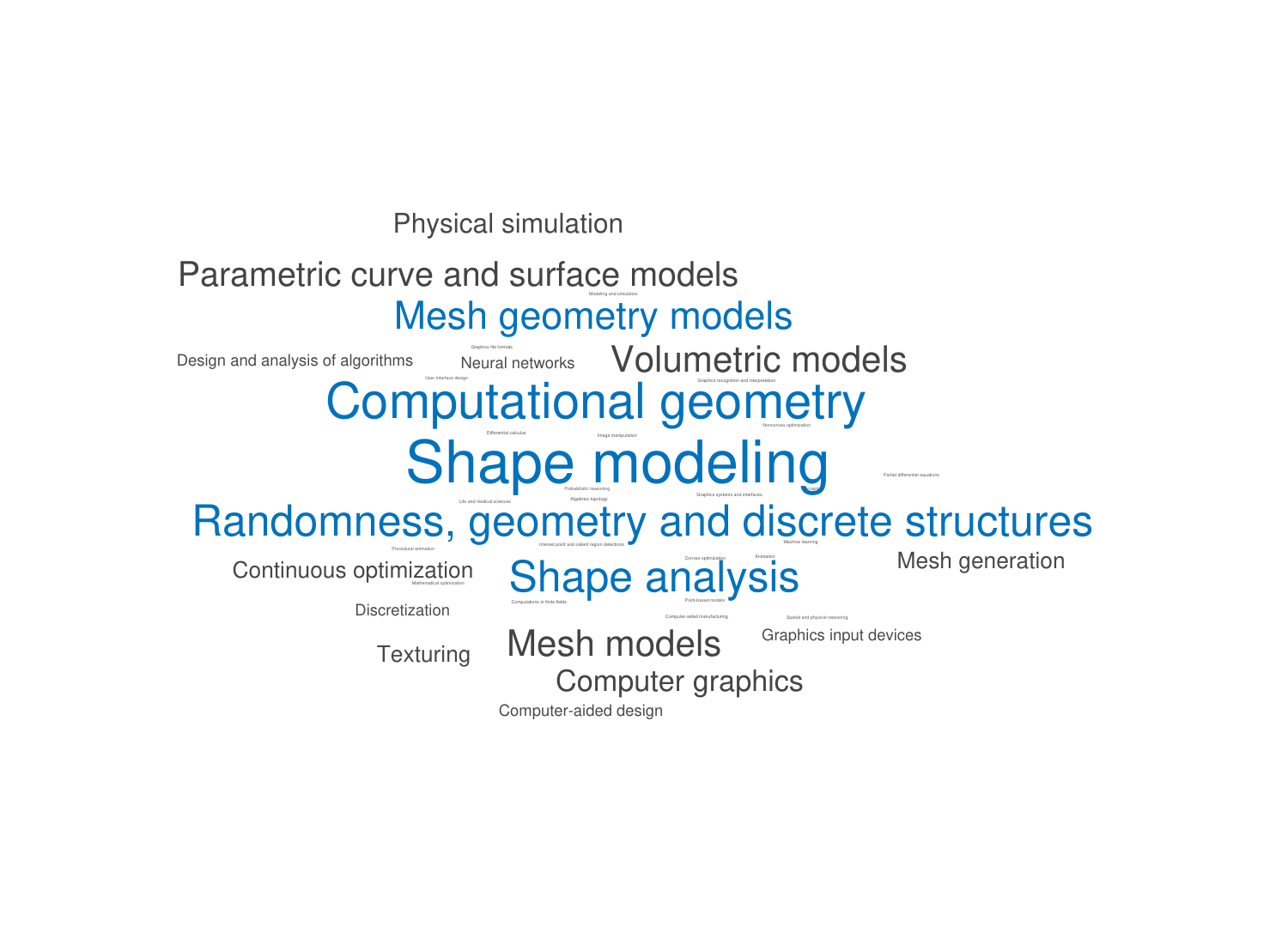}
 \caption{Geometry}
\end{subfigure}%
\begin{subfigure}{.45\columnwidth}
 \centering
 \includegraphics[width=\columnwidth]{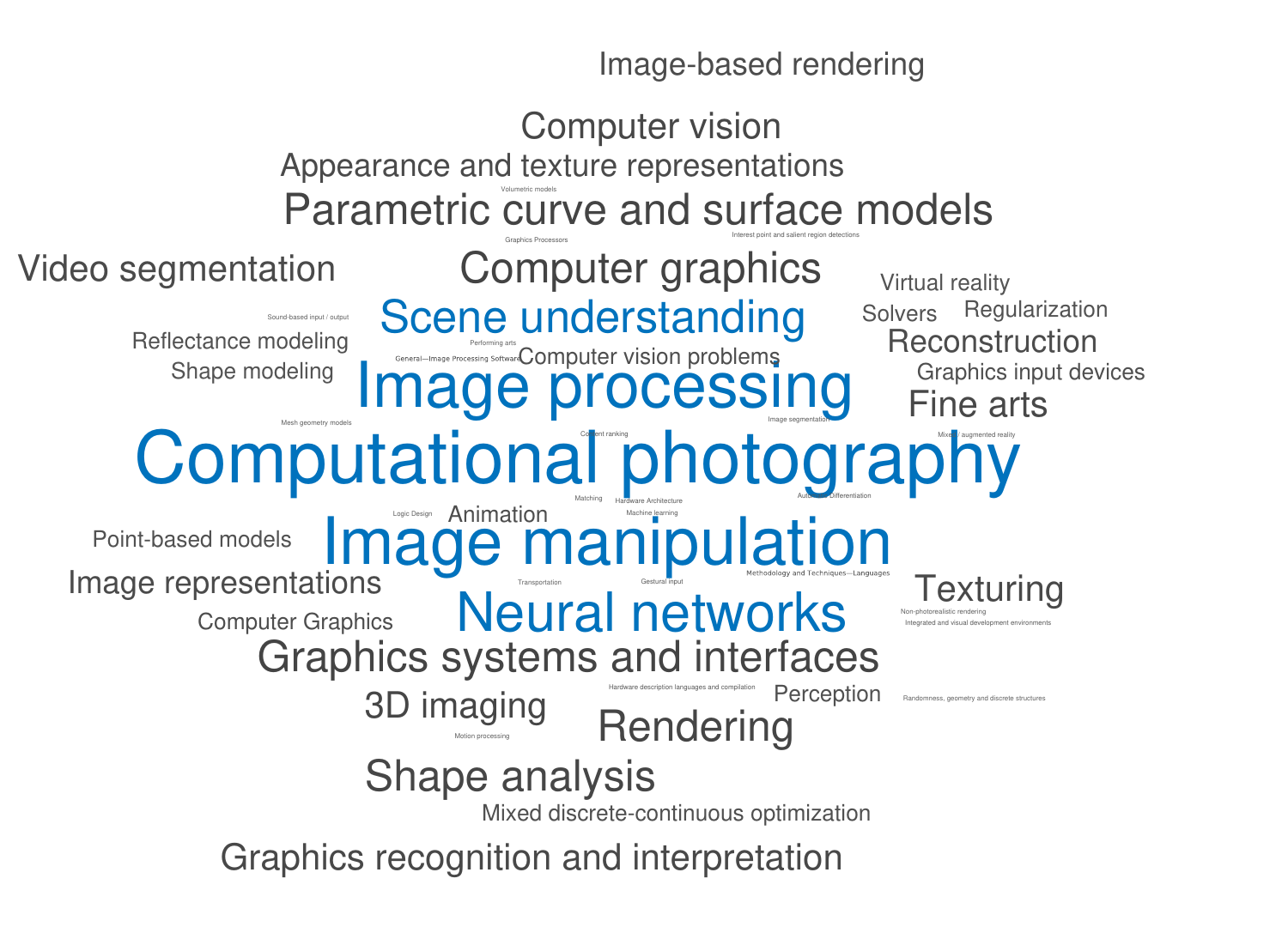}
 \caption{Image}
\end{subfigure}%
\newline
\begin{subfigure}{.45\columnwidth}
 \centering
 \includegraphics[width=\columnwidth]{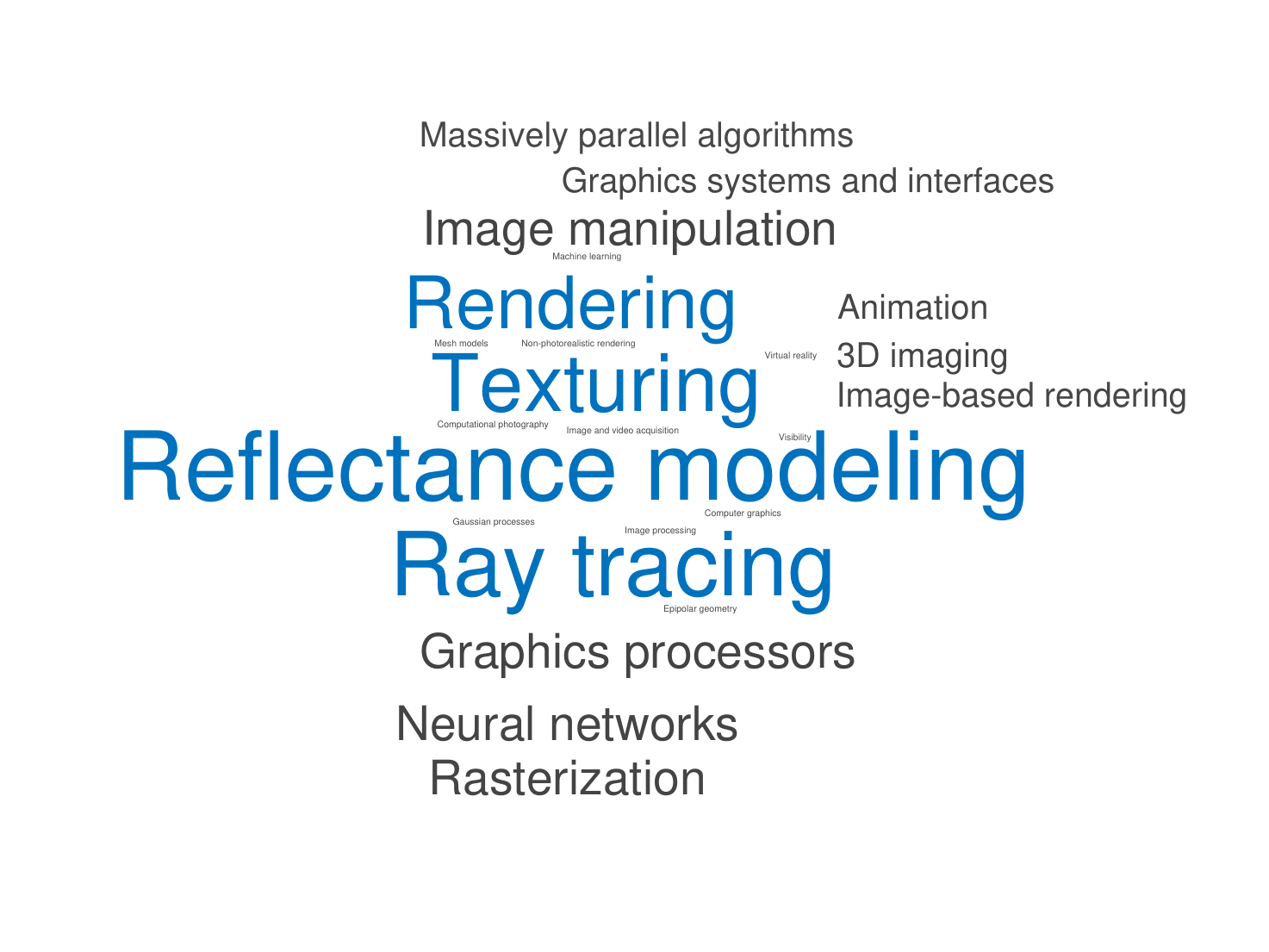}
 \caption{Rendering}
\end{subfigure}%
\begin{subfigure}{.45\columnwidth}
 \centering
 \includegraphics[width=\columnwidth]{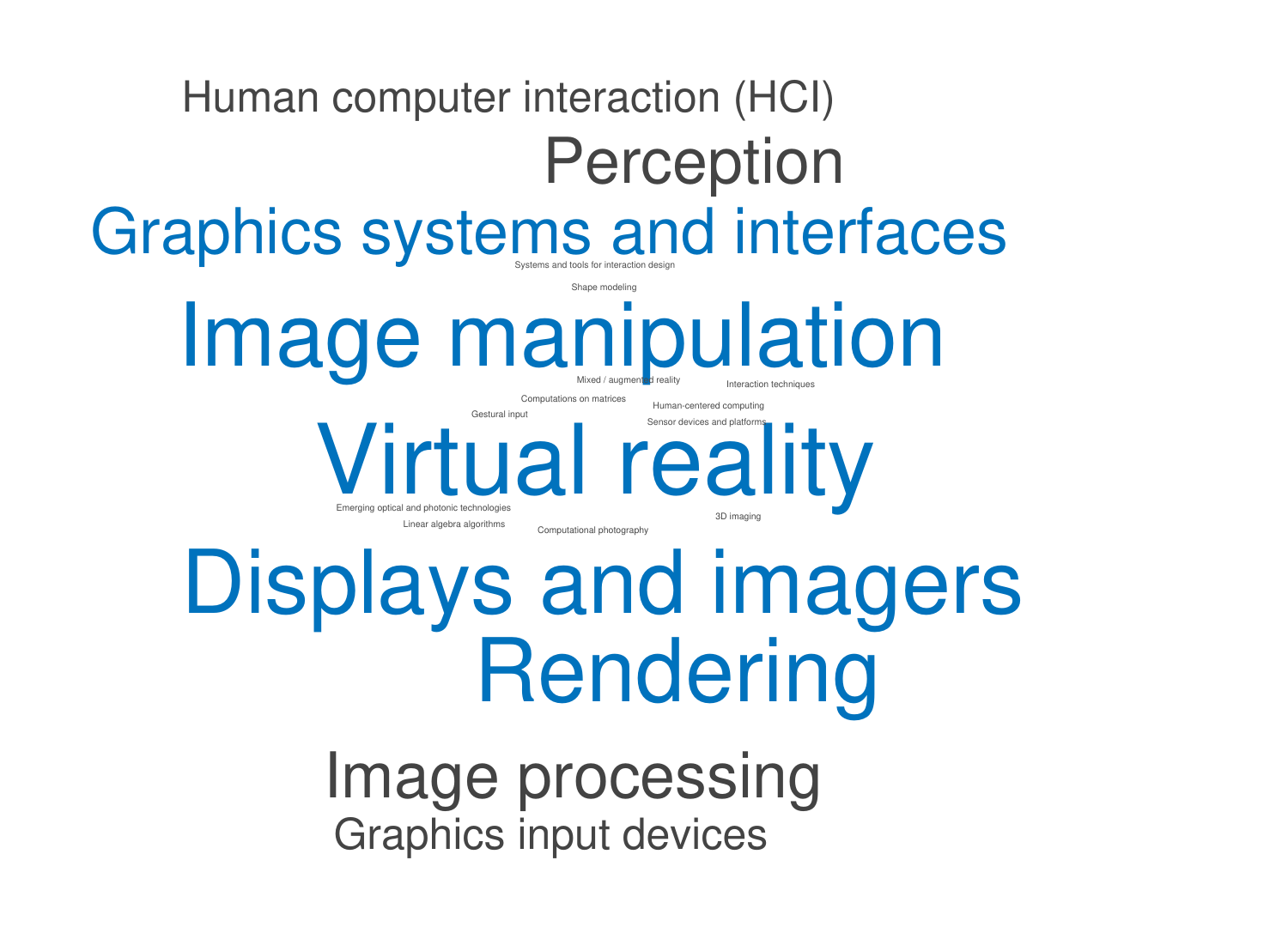}
 \caption{Virtual Reality}
\end{subfigure}%
\caption{Distribution of the ACM keywords per topic. The font size reflects the number of papers associated with a keyword.}
\vspace{-0.3cm}
\label{fig:acm-keywords-cloud}
\end{figure}

\textbf{Identifying and factual information}. This includes the paper
name and DOI, ACM keywords, pdf, project and code or binaries URLs if
they have been found, as well as information indicating if authors are from
the industry, academia, or unaffiliated, for further analysis. For
papers, we include information as whether they can be found on arXiv or other \emph{Open Archive Initiative} providers we  may
have found,
in open access on the ACM Digital Library, or by other means such as institutional web pages.
Aside from ACM keywords, we further categorize papers into 6 broad topics related to computer graphics, and we also keep track of whether they relate to neural networks. We defined these topics as:

\begin{itemize}
\item\emph{Rendering}. This includes simulating light transport, real-time rendering, sampling, reflectance capture, data-structures for intersections, and non-photorealistic rendering.
\item\emph{Animation and simulation}. This includes character animation, motion capture and rigging/skinning, cinematography/camera path planning, deformable models as well as fluid, cloth, hair or sound simulation, including geometric or topology problems related to these subjects.
\item\emph{Geometry}. This includes geometry processing and modeling, for
  point-based, voxel-based and mesh-based geometries, as well as
  topology, mapping, vector fields and shape collection
  analysis. We also include image-based modeling.
\item\emph{Images}. This includes image and video processing, as well as texture synthesis and editing, image segmentation, drawing, sketching and illustration, intrinsic decomposition or computational photography. We also included here image-based rendering, which relies more on image techniques than rendering.
\item\emph{Virtual Reality}. This category includes virtual and augmented reality, 3d displays, and interactions.
\item\emph{Fabrication}. This includes 3d printing, knitting or caustic design.
\end{itemize}
We strive to classify each paper into a single category to simplify analyses. Both these categories and paper assignments to these categories can be largely debated. While they may be prone to errors at the individual level, they still provide meaningful insight when seen as statistical aggregates. These categories were used in our analysis instead of ACM keywords for several reasons: first, we counted more than 127 different ACM keywords which would make overspecialized categories. The hierarchical nature of this taxonomy also makes the analysis more complicated. In Fig.~\ref{fig:acm-keywords-cloud} we show the distribution of ACM keywords of papers involved in each of our categories. Interestingly, this visualization exacerbates the lack of ACM keywords dedicated to fabrication despite the increasing popularity of this topic.

Information about code {includes} code license, presence of documentation, \texttt{readme} files and explicit mention of the code authors (who usually are a subset of the paper authors), as well as build mechanism (\texttt{Makefile}, \texttt{CMakeLists}, \texttt{SCons}, IDE projects, or other types of scripts), and lists of dependencies. We notably indicate whether library or software dependencies are open source (e.g., \texttt{Eigen}, \texttt{OpenCV}), closed source but free at least for research purpose (e.g., \texttt{mosek}, \texttt{CUDA} or Intel \texttt{MKL}), or closed source and paying even for research purpose (e.g., \texttt{Matlab}). Similarly, we ask whether the code depends on data other than examples or input data (e.g., training data or neural network description files) and their license.

One of our key contributions is that we report the undocumented steps required to make the code run -- from bug fixes to dependency installation procedures. We believe this information is valuable to the community as these steps are often independently found by students relying on these codes sometimes after significant effort.

\textbf{Subjective judgments on replicability}. For papers without published code, this includes information as to whether the paper contains explicit algorithms and how much effort is deemed required to implement them (on a scale of 1 to 5). For algorithms requiring little reimplementation effort (with a score of 5) -- typically for short shaders or short self-contained algorithms -- this can give an indication as to why releasing the code was judged unnecessary.
For papers containing code, we evaluate  how difficult it was to replicate results through a number of questions on a scale of 1 to 5. This includes the difficulty to find and install dependencies, to configure and build the project, to fix bugs, to adapt the code to other contexts, and how much we could replicate the results shown in the paper. We strived to remain flexible in the replicability score: often, the exact input data were not provided but the algorithms produced satisfactory results that are  qualitatively close to those published on different data, or algorithms relied on random generators (e.g., for neural network initializations) that do not produce repeatable number sequences and results. Contrary to existing replicability initiatives, we did not penalize these issues, and this did not prevent high replicability scores.

We shared the task of evaluating these \npapers~submissions across 4
full-time tenured researchers (authors of the paper), largely
experienced in programming and running complex computer graphics
systems.  Reasonable efforts were made to find and compile the
provided code, including retrieving outdated links from the WayBack
Machine~\cite{tofel2007wayback}, recreating missing \texttt{Makefiles}, debugging, trying on
  multiple OS (compiling was tested on Windows 10, Debian Buster, Ubuntu 18.04 and
  19.10 and MacOS 10.15\footnote{Ubuntu 14.04 and Windows 2012 virtual machines for
      very specific tests.}), or adapting the code to match libraries
  having evolved.
Efforts to adapt the code to evolved libraries, compilers or languages are due to practical reasons: it is sometimes impractical to rely on old Visual Studio 2010 precompiled libraries when only having access to a newer version, or to rely on \texttt{TensorFlow} 1.4.0 requiring downgrading \texttt{CUDA} drivers to version 8 for the sole purpose of having a single code run.
We chose to avoid contacting authors for clarifications, instructions or to report bug fixes to protect anonymity. We also added the GitHub projects to Software Heritage~\cite{dicosmo17} when they were not already archived and gave the link to the Software Heritage entry in our online tool.

\section{Data Exploration}
We provide the data collected during our review as a JSON file, available as supplementary material.
Each JSON entry describes the properties of a paper (e.g., author list, project page, ACM keywords, topics) and its replicability results (e.g., scores, replicability instructions).
All the indicators and statistics given in this paper are computed from this data, and we provide in supplementary materials all the scripts required to replicate our analysis.

\begin{figure*}[!tb]
\includegraphics[width=.65\columnwidth]{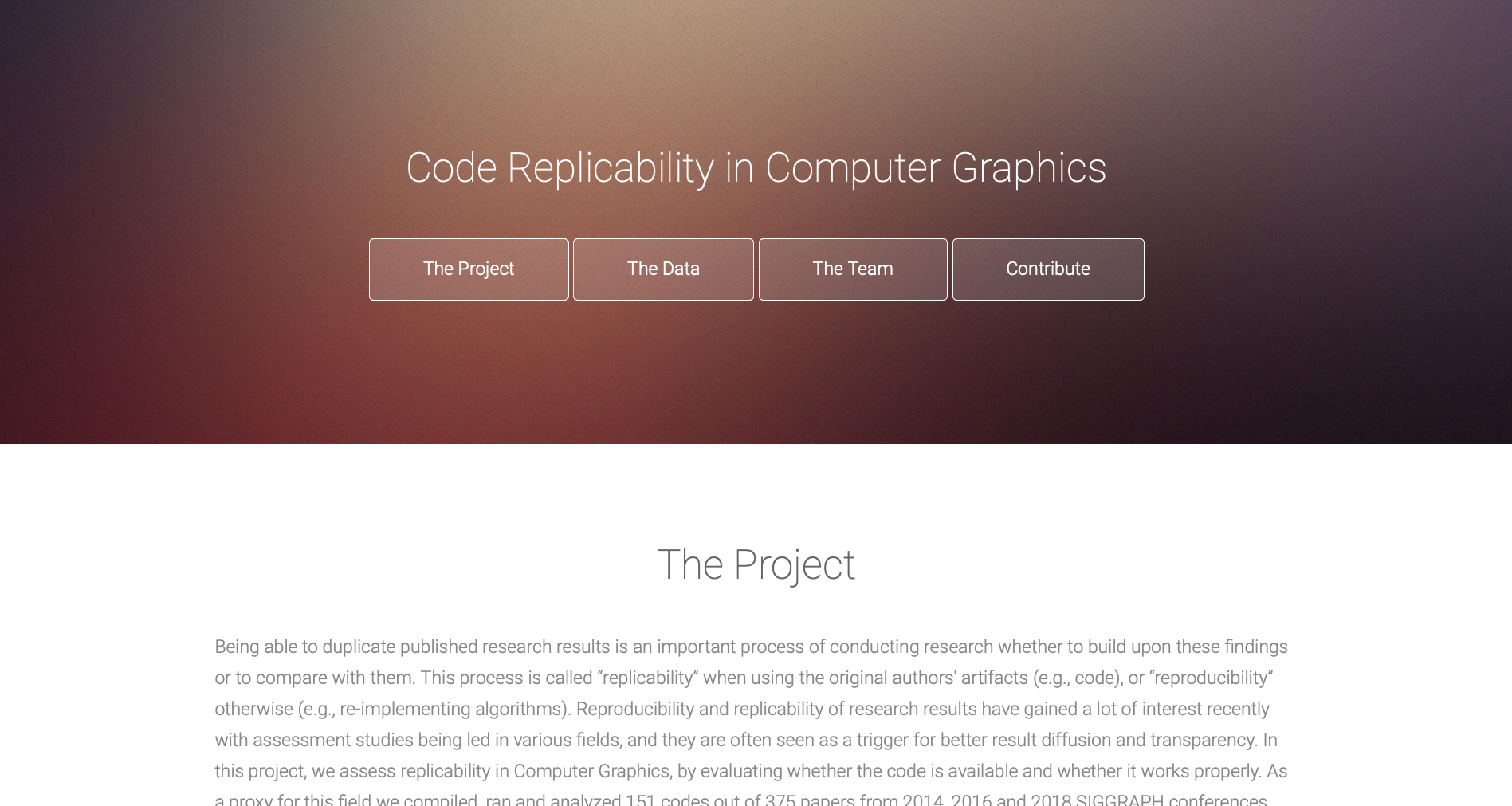}\hspace{.2cm}
\includegraphics[width=.65\columnwidth]{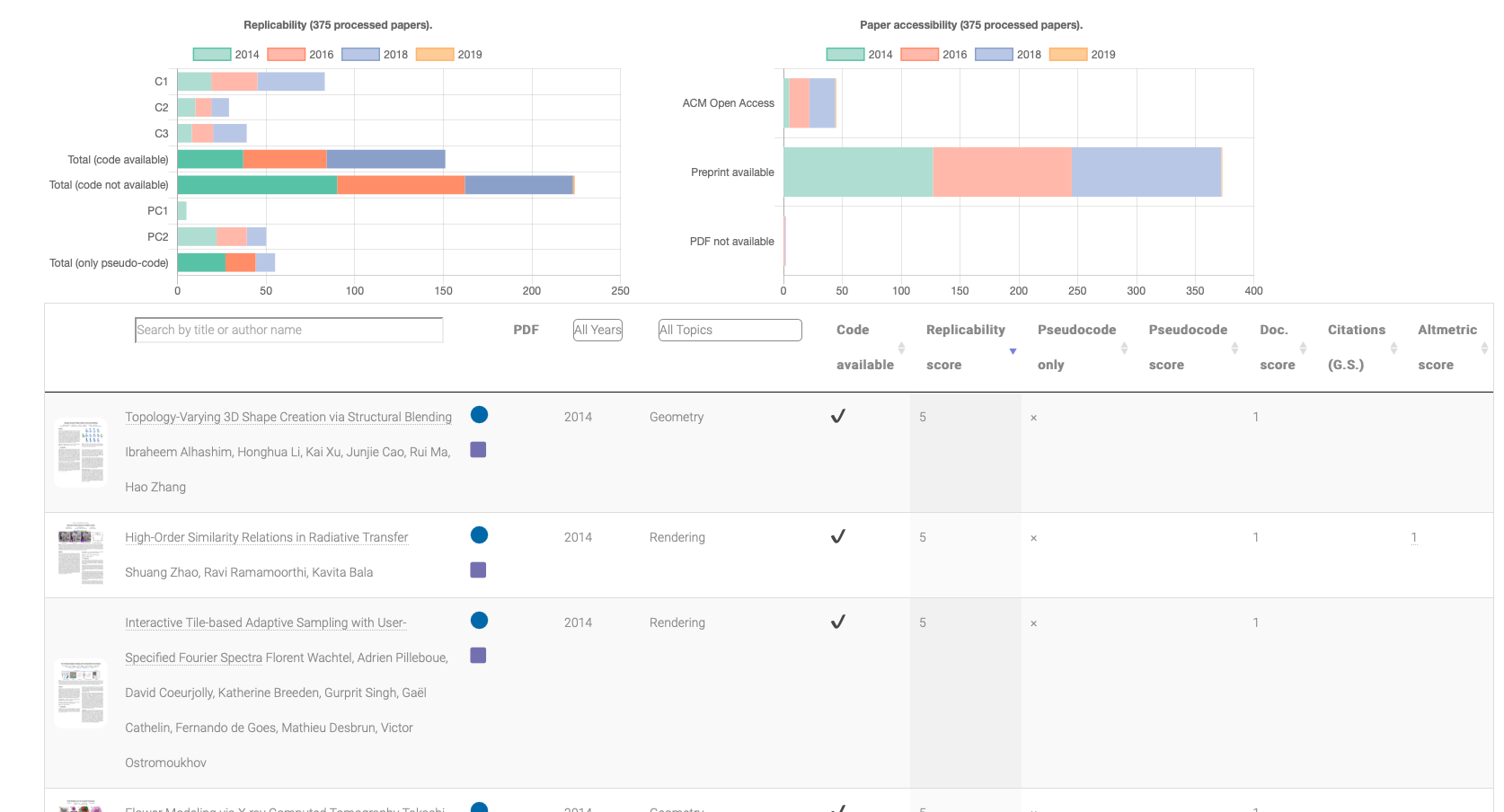}\hspace{.2cm} 
\includegraphics[width=.65\columnwidth]{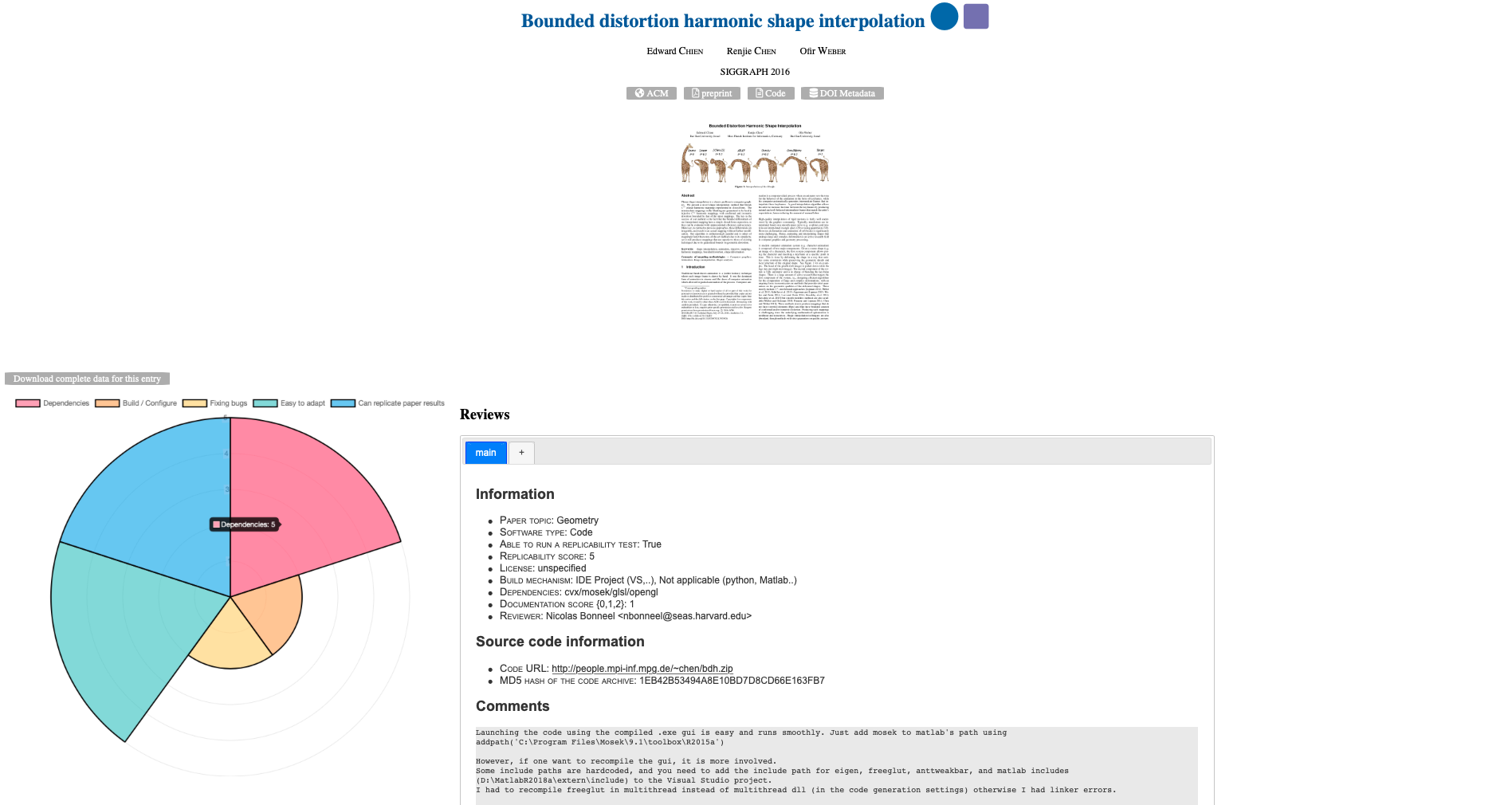}
\caption{We designed a web interface to explore our collected data
  allowing to see individual paper replicability and build
  instructions, available at \url{https://replicability.graphics}.}
  \vspace{-0.3cm}
  \label{fig:web}
\end{figure*}

We facilitate data exploration by providing an intuitive web interface available at \url{https://replicability.graphics} (see Fig.~\ref{fig:web}) to visualize collected data.
This interface allows two types of exploration, either the whole dataset or per paper.


\paragraph{Dataset exploration}
Our dataset exploration tool is split into two components: a table
listing the reviewed papers, and two graphs showing statistics about the
table content.  A first graph displays the distribution of papers
with respect to the code/pseudocode availability, and their replicability
score.  A second graph shows papers availability, either as
ACM Open Access or as a preprint provided by the authors.  The
interactive table allows to filter the dataset by {the} author name, paper
title, publication year and/or by topic, and to update the graphs
according to the selection. It is also possible to sort the paper by
their properties and in particular their replicability score or a
documentation score between $0$ and $2$ ($0$: no documentation, $2$:
 exhaustive documentation).  Each paper is associated with a
dedicated webpage accessible directly from the table.

\paragraph{Per-paper exploration}
The paper webpage gives a direct access to the information extracted from the JSON file.
It includes the links to {resources} available online (Digital ACM library, preprint, code), several {information} (e.g., paper topic, nature of the {artifact}, list of the dependencies) and a breakdown of the replicability experiment when code was available (scores and comments).
In addition, the paper webpage gives the Altmetric Attention Score\footnote{\url{https://www.altmetric.com/}} and links to the Altmetric webpage of the paper if available. This score measures the overall attention a paper has received, including on social networks, which differs from academic citation scores.
The comment section mostly covers the steps that the reviewer had to
follow in order to try to replicate the paper, which includes details
about {the} dependencies management and updates, bug fixes or code modifications. We expose the exact revision number (for git projects) or MD5 hash of the archive file (for direct download) of codes that relate to the comments.
The website allows for commenting scores and instructions, both as a user and as a paper author, as well as adding new entries, for adding new or updated codes.

\section{Results and Analysis}

This section analyzes both objective and subjective metrics. All reported p-values were adjusted for multiple comparisons using the false discovery rate control procedure proposed by~\citet{benjamini1995controlling}.

\subsection{Objective analysis}
\label{sec:objectiveanalysis}

\textbf{Availability of papers.} Papers are overall available. Over all
\npapers~papers, only two are available only on the ACM
Digital  Library. Notably, ACM provides free access to all SIGGRAPH and SIGGRAPH Asia 
proceedings though it is little advertised~\cite{acmlib}. Also, \narxiv~are
available as preprints on arXiv (9 only on arXiv), 17~on HAL (7
  only on HAL)\footnote{Some references or preprints may also be
      available on other OAI providers thanks to database
      interconnections or local initiatives, we only report here the most
      significant ones found by this study.},
\nopenaccess~benefit from the ACM Open Access policy --the other
papers being available as preprints at least on the authors website or
other paper repositories. 

\textbf{Availability of code.} Software packages were available for \ncodes~papers, which consist of \nsourcecode~papers for which source code was provided plus \nbinaries~papers for which no source code was provided but instead compiled software was provided. For the rest of the analysis, we considered both compiled and open source software combined. While open source research codes allow for adaptation, making it easier to build upon them, and are thus ideal, binary software at least allows for effortless method comparisons. Nevertheless, among these software packages, we could not run \nsoftissues~of them due to technical issues preventing the codes to compile or run, and \nhardissues~of them due to lack of dedicated hardware (see Sec.~\ref{sec:limitations}).
Among these \nsourcecode~codes, \nunspeclicence~do not have  license information, which could notably prevent code dissemination in the industry, and \nodoc~do not come with any documentation nor build instructions.

\begin{figure}[!tb]
\includegraphics[width=0.47\columnwidth]{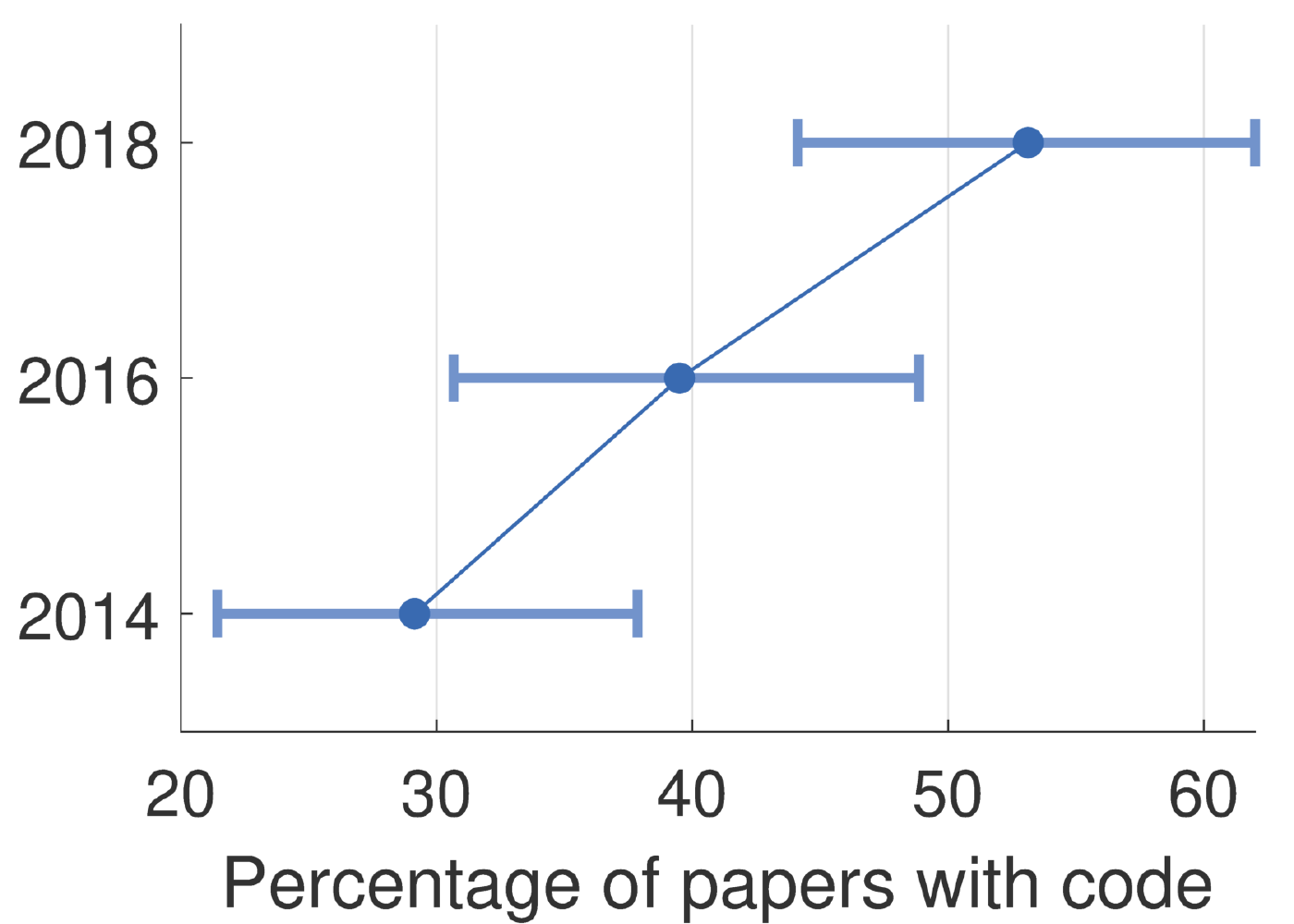}
\includegraphics[width=0.47\columnwidth]{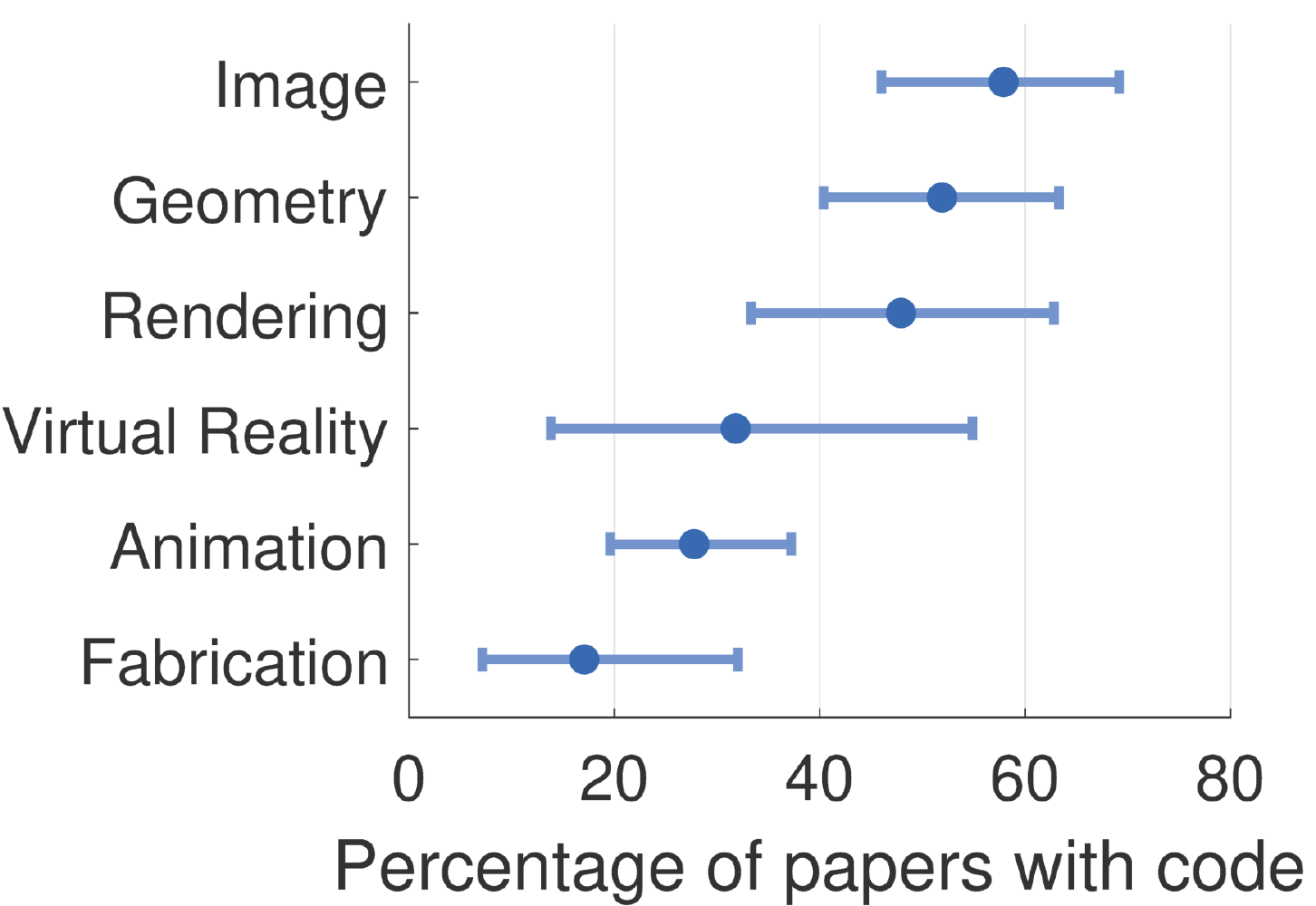}
\caption{We compute the percentage of papers that include either code or binaries as a function of years and topic. We also show Clopper-Pearson 95\% confidence intervals.}
\vspace{-0.4cm}
\label{fig:codeyear}
\end{figure}

We perform $\chi^2$ tests to analyze trends in code sharing.
Overall, codes or binaries could be found for \ncodesIV~papers out of \npapersIV~(\pccodesIV\%) in 2014,  \ncodesVI~out of \npapersVI~(\pccodesVI\%)~in 2016, \ncodesVIII~papers out of \npapersVIII~(\pccodesVIII\%) in 2018.
This increase is statistically  significant between 2014 and 2018 ($p = \pcodesIVVIII$), though not between 2014 and 2016 ($p = \pcodesIVVI$) nor between 2016 and 2018 ($p = \pcodesVIVIII$). This trend is similar to that observed in artificial intelligence~\cite{hutson2018artificial,gundersen2018state}. Results can be seen in Fig.~\ref{fig:codeyear}. In two cases, we had to retrieve the code from the WayBack Machine~\cite{tofel2007wayback} due to expired URLs. Further analysis per topic shows vastly different practices, with \pccodesFab\% of papers sharing code for \emph{Fabrication}, \pccodesAni\% for \emph{Animation}, \pccodesVR\% for \emph{Virtual Reality}, \pccodesRen\% for \emph{Rendering}, \pccodesGeo\% for \emph{Geometry} and \pccodesIma\% for \emph{Images} (Fig.~\ref{fig:codeyear}).

We also analyzed the involvement of at least one author from the industry on the release of codes or binaries. We found that overall, papers involving the industry provided code or binaries \pccodesindus\% of the times, while this was the case for \pccodesacad\% of purely academic papers -- a difference that is significant ($p = \pcodesindus$). This could be explained by strict rules imposed by employers, understandably worried about industrial consequences of sharing a code.

Given the sheer amount of deep learning codes available online, we hypothesized that deep learning-related papers were more likely to share code. We tested this hypothesis on our dataset,
but we found that they provided code only \fcodesdeep\% of the times (\ncodesdeep~out of \npapersdeep), while this
was the case \fcodesnodeep\% of the times for non-deep papers (\ncodesnodeep~out of \npapersnotdeep) -- a non-significant difference ($p=\pcodesdeepn$).

\begin{figure}[!tb]
\includegraphics[width=0.7\columnwidth]{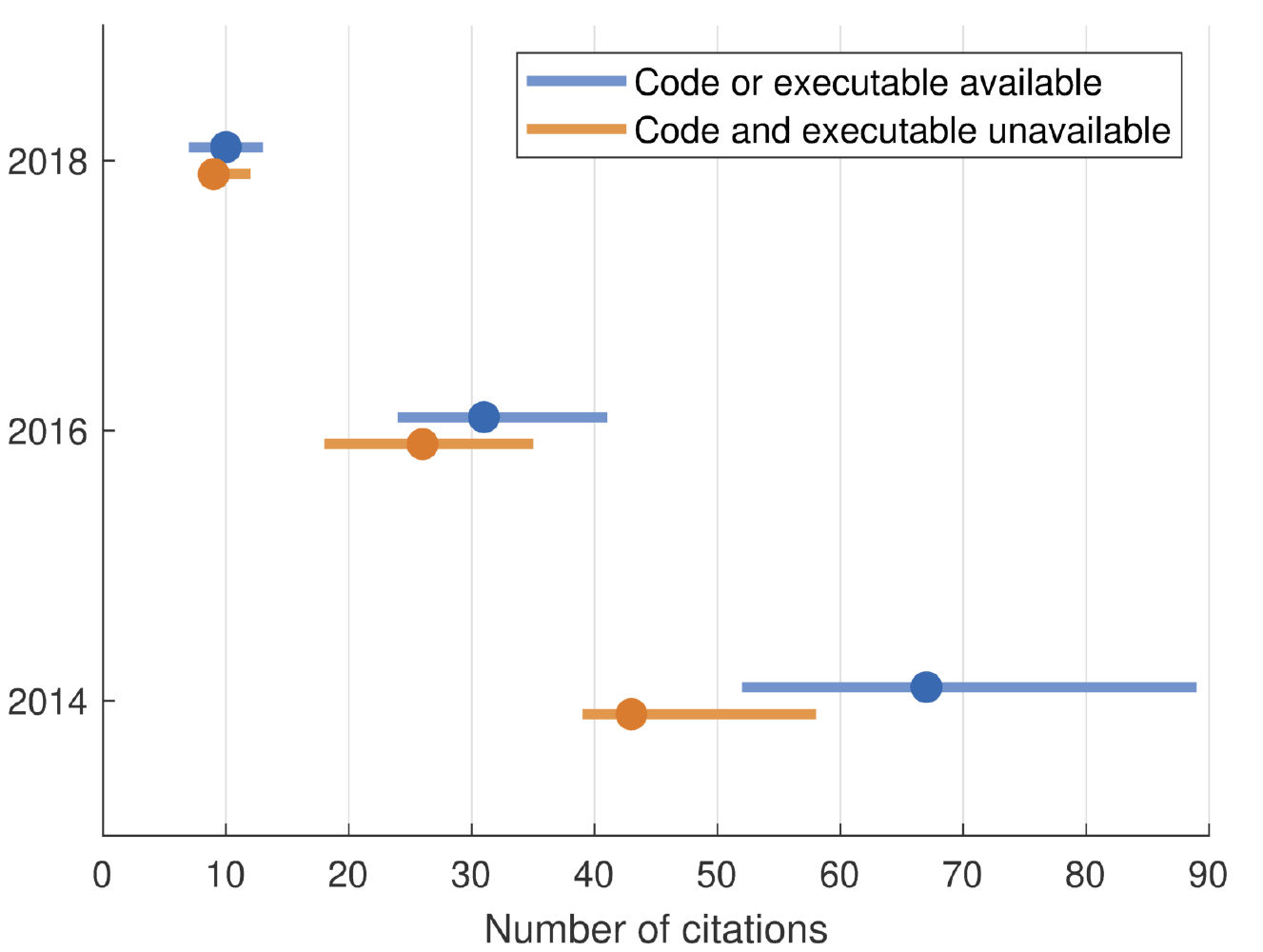}
\caption{We compute the median number of citations and its 95\% confidence intervals for papers sharing code (or executable) and for papers not sharing code nor executable.}
\vspace{-0.3cm}
\label{fig:codecit}
\end{figure}

We finally found that, in the long term, sharing code results in
higher citations, with a median citation count of up to \citcodeIV~ in 2014
for papers sharing code compared to \citnocodeIV~for papers not
sharing code~(see Fig.~\ref{fig:codecit}). A Mann-Whitney U-test gives
this difference significant ($p=\pcitcodenocodeIV$). This observation
is similar to that observed in image processing~\cite{Vandewalle19}
though the effect is less pronounced (they observed a doubling of
citation rates). Few additional information {is} given in Table \ref{tab:letableau}.

\subsection{Subjective analysis}

As replicability scores are subjective, we first perform an analysis of variance (ANOVA), despite some limitations here (see Norman~\shortcite{norman2010likert} for a full discussion), aimed at determining two things: is there a dependence on the reviewer of the code on replicability scores? And, does the year influence replicability (as it would seem that older non-maintained codes are harder to replicate)? The ANOVA is performed on the replicability score, taking only papers with codes for which compiling was successful, and with two factors: reviewer and year.
The answer to both questions seems negative (resp. $p=\preplreviewer$ and $p=\preplyear$).

To make the codes run, we had to modify source codes in \nmodifiedcodes~out of \ncodes~codes. These code alterations were deemed difficult (``easy to fix bugs'' score $\leq$ 2 out of 5) for \nhardmodifiedcodes~codes. The time spent to make codes run, including time to debug and compile dependencies was longer than 100 minutes for \nlongcodes~codes.

In the years covered by this study, we found a total of 5 papers with a Replicability Stamp from the Graphics Replicability Stamp Initiative~\cite{grsi2016}. While this number is too low to derive meaningful statistics, one can note that out of these 5 papers, 4 get the maximum score for results replication. This could be expected because this initiative ensures that a script for each single result shown in the paper is provided. A limitation is that these scripts are only guaranteed to work at the time when the stamp is granted --a limitation shared by the present study.

\begin{table}
  \caption{Additional quantitative data from our study.\label{tab:letableau}}
  \begin{tabular}{l|c|c|c|r}
    &2014 & 2016 & 2018 & Total\\ \hline 
    Nb of papers  & \npapersIV & \npapersVI  & \npapersVIII & \npapers\\
    Nb of papers with codes &  \ncodesIV&  \ncodesVI&  \ncodesVIII& \ncodes\\\hline\hline
    Nb of codes without license  & \nolicIV& \nolicVI& \nolicVIII&\nolicTot\\
    Nb of codes with doc. score 0  & \doczeroIV& \doczeroVI& \doczeroVIII&\doczeroTot\\
    Nb of codes flagged as ``Deep''  & \deepIV& \deepVI& \deepVIII&\deepTot\\
    Nb of codes with associated data  & \dataIV& \dataVI& \dataVIII& \dataTot\\
    \hline
  \end{tabular}
  \vspace{-0.2cm}
\end{table}

\section{Limitations}
\label{sec:limitations}

Our analysis has a number of limitations. First, the data we collected
may only be partly reliable.
 While we spent reasonable efforts to find, run and
compile codes, it is possible that we missed codes, or that additional efforts or contacting
the authors for clarifications or to report bugs would result in
different outcome for a few papers. Similarly, we could not fully
evaluate codes that depend on specific hardware (such as spatial
light modulators, microcontrollers, Hall effect sensors etc.) for 4
papers.  Our analysis focused on assessing the codes provided by the
authors which only assesses replicability but not reproducibility:
there are instances for which papers were successfully reimplemented by
other teams, which falls out of our analysis scope. It could also be
expected that certain codes could be available upon request ; in fact,
in a few cases, the provided code relied on data only available upon
request, which we did not assess.

Second, the codes we found and assessed may have evolved after the paper has been published, which we cannot control. Similarly, the published code could be a cleaned-up version of the original code, or even a full reimplementation.

Third, our focus on SIGGRAPH could hide a more negative picture of the entire field. We believe that the exposure SIGGRAPH probably gives  biases our results, with a tendency to find more codes here than in smaller venues. It would be an interesting future work to compare replicability across computer graphics venues.

\section{Recommendations}

Our replicability evaluation led us to identify a number of issues.
First, the number of dependencies was often correlated with the difficulty to compile -- especially on 
Windows. Precompiled libraries were sometimes provided for compilers that became outdated, or some dependencies were no longer supported on recent hardware or OS.  The lack of precise dependencies version number was another important issue we faced. Package managers for Python such as \emph{pip}  or \emph{conda}
  evolve and default to different library versions, and build instructions or installation scripts did not directly work with these new versions.
  Lack of instructions for running software raised important frustration: default parameters 
  were sometimes not provided, command line parameters not described, or results output as numerical values  
  in the console or written to files of undocumented format with no
  clear ways to use. In one case we had to develop a 
  software for reading the output file and displaying results. Similarly, input data were not always provided, or sometimes only provided upon request. Finally, in some cases, the software implemented only 
  part of the method, producing results that did not match the quality of the results shown in the paper (e.g., missing 
  post-processing, or only implementing the most important parts of the paper).
  
This leads us to issue several recommendations for
the research community in Computer Graphics to promote their research
work through the code replicability point of view.

\paragraph{\textbf{For the  authors.}} Sharing the code or any
artifact to help the replicability of the paper results is a good way
to spread the contributions of the paper, as shown in terms of
citation numbers in Sec.~\ref{sec:objectiveanalysis} and independently
by Vandewalle~\shortcite{Vandewalle19}.  When shipping the code as a
supplementary material for the paper, several concerns must be
addressed: Accessibility of the source code ({e.g.,} using the
Software Heritage archive when the article is accepted) ;
Replicability of the build process specifying the exact versions of
the software or libraries that must be used to build and execute the
code (for instance using container/virtualization services
--\texttt{docker\footnote{\url{https://www.docker.com}}}-- or
package/configurations managements services
--\texttt{anaconda\footnote{\url{https://anaconda.org}},
  Nix\footnote{\url{https://nixos.org}}} etc.); Clarity of the
source code as a knowledge source ({e.g.,} through technical
documentation and comments in the code); and, finally, tractability of
the coding process (authorship, clear licensing etc.).   
  Extra care should be given to
codes that depend on rapidly evolving libraries. This is particularly
the case of deep learning libraries (\texttt{TensorFlow},
\texttt{Pytorch}, \texttt{Caffe} etc.). As an example, several syntax
changes occurred in \texttt{pytorch} over the past few years and caffe
appears {not to} be maintained anymore (e.g., pre-built binaries are
provided up to Visual Studio 2015, and the last commit on Github was
in March 2019) ; Python 2.7 is not maintained anymore as of January
1st, 2020.  
We recommend limiting the number of dependencies when possible -- e.g., avoiding to 
depend on large libraries for the sole purpose of loading image files -- possibly 
shipping dependencies with the source code (with integration into the project build framework).
Similarly, deep learning codes can require up to several
days of training: sharing pre-trained models together with the
training routines is a good way to ensure replicability.

\paragraph{\textbf{For the conference program chairs.}} Not all research papers need to be involved in a source code replicability effort. A paper presenting a mathematical proof of the asymptotic variance of some sampler is intrinsically reproducible and does not need source code. On the contrary, for research papers for which it would make sense, the source code should be
considered as a valuable artifact when evaluating a submission. This
can be emphasized and encouraged in the guidelines and submission
forms. Asking the reviewers to check the replicability of the article
through the provided code is an ultimate goal for targeting
replicability but it may not be sustainable for the entire
community. Intermediate action could be to communicate about the
importance of paper replicability and to allow authors to submit their
(anonymous) code and data with appropriate entries in the submission
system. Furthermore, we advocate for a specific deadline for the
  submission of the code and data materials, \emph{e.g.} one week
  after the paper deadline. The objective would be to let additional time for authors to 
  sanitize their code and encourage its publication, without interfering with the intense paper polishing process happening right before the paper deadline, nor with the reviewing process, since reviewers would only wait for a short amount of time before getting the code.

\paragraph{\textbf{For the  publishers.}} Publishers already offer the
possibility to attach supplementary materials to published papers
(videos, source code, data\ldots). Beside videos, other types of
supplementary documents are not clearly identifiable. We would
recommend to clearly tag (and promote on the publisher library)
supplementary data that correspond to source codes.
Independent platforms, such as Software Heritage~\cite{dicosmo17}, permit archiving and attach unique identifiers to source codes as well as a timestamp. Publishers could easily rely on these platforms to link to the source code of a paper.

\section{Conclusion}

Our in-depth study of three years of ACM SIGGRAPH conference papers showed a clear increase in replicability as authors are increasingly sharing their codes and data. Our study also showed that sharing code was correlated with a higher impact, measured in terms of citation numbers. We developed a website which aims at helping practitioners run existing codes on current hardware and software generations, with build instructions for \ncodes~codes we found online when we could run them.
Contrary to existing replicability stamps, our replicability scores are non-binary but on a 1-to-5 integer scale, less strict in the sense that results sufficiently close those shown in the paper on different data were deemed appropriate, but sometimes inconsistent with these stamps when software could not be run \textit{anymore} on current hardware and software generations.
In the future, we hope to see interactions with these replicability stamp initiatives for which we share the common goal of spreading open research.

\section{Acknowledgments}
We thank Roberto Di Cosmo for insightful discussions and the reviewers for their constructive feedback. This work was funded in part by ANR-16-CE23-0009 (ROOT), ANR-16-CE33-0026 (CALiTrOp), ANR-15-CE40-0006 (CoMeDiC) and ANR-16-CE38-0009 (e-ROMA).

\bibliographystyle{ACM-Reference-Format}
\bibliography{paper}

\end{document}